\documentclass[12pt]{article}
\usepackage{amsmath}
\usepackage{graphicx,psfrag,epsf}
\usepackage{enumerate}
\usepackage{natbib}
\usepackage{textcomp}
\usepackage[hyphens]{url} 
\usepackage{hyperref}
\providecommand{\tightlist}{%
  \setlength{\itemsep}{0pt}\setlength{\parskip}{0pt}}

\newcommand{\blind}{0}

\usepackage{color}
\usepackage{fancyvrb}

\DefineVerbatimEnvironment{Highlighting}{Verbatim}{commandchars=\\\{\}}
\usepackage{framed}
\definecolor{shadecolor}{RGB}{248,248,248}
\newenvironment{Shaded}{\begin{snugshade}}{\end{snugshade}}

\newcommand{\AttributeTok}[1]{\textcolor[rgb]{0.77,0.63,0.00}{#1}}

\newcommand{\CommentTok}[1]{\textcolor[rgb]{0.56,0.35,0.01}{\textit{#1}}}

\newcommand{\ControlFlowTok}[1]{\textcolor[rgb]{0.13,0.29,0.53}{\textbf{#1}}}

\newcommand{\DecValTok}[1]{\textcolor[rgb]{0.00,0.00,0.81}{#1}}
\newcommand{\DocumentationTok}[1]{\textcolor[rgb]{0.56,0.35,0.01}{\textbf{\textit{#1}}}}

\newcommand{\FunctionTok}[1]{\textcolor[rgb]{0.00,0.00,0.00}{#1}}

\newcommand{\InformationTok}[1]{\textcolor[rgb]{0.56,0.35,0.01}{\textbf{\textit{#1}}}}

\newcommand{\NormalTok}[1]{#1}

\newcommand{\OtherTok}[1]{\textcolor[rgb]{0.56,0.35,0.01}{#1}}

\newcommand{\SpecialCharTok}[1]{\textcolor[rgb]{0.00,0.00,0.00}{#1}}

\newcommand{\StringTok}[1]{\textcolor[rgb]{0.31,0.60,0.02}{#1}}

\usepackage{booktabs}
\usepackage{longtable}
\usepackage{array}
\usepackage{multirow}
\usepackage{wrapfig}
\usepackage{float}
\usepackage{colortbl}
\usepackage{pdflscape}
\usepackage{tabu}
\usepackage{threeparttable}
\usepackage{threeparttablex}
\usepackage[normalem]{ulem}
\usepackage{makecell}
\usepackage{xcolor}

\begin{document}

\def\spacingset#1{\renewcommand{\baselinestretch}%
{#1}\small\normalsize} \spacingset{1}


\if0\blind
{
  \title{\bf Framework for Accessible and Inclusive Teaching Materials for Statistics and Data Science Courses}

  \author{
        Mine Dogucu \\
    Department of Statistics, University of California Irvine\\
     and \\     Alicia A. Johnson \\
    Department of Mathematics, Statistics, Computer Science, Macalester College\\
     and \\     Miles Ott \\
    The Janssen Pharmaceutical Companies of Johnson \& Johnson\\
      }
  \maketitle
} \fi

\if1\blind
{
  \bigskip
  \bigskip
  \bigskip
  \begin{center}
    {\LARGE\bf Framework for Accessible and Inclusive Teaching Materials for Statistics and Data Science Courses}
  \end{center}
  \medskip
} \fi

\bigskip
\begin{abstract}
Despite rapid growth in the data science workforce, people of color, women, those with disabilities, and others remain underrepresented in, underserved by, and sometimes excluded from the field. This pattern prevents equal opportunity for individuals, while also creating products and policies that perpetuate inequality. Thus, for statistics and data science educators of the next generation, accessibility and inclusion should be of utmost importance in our programs and courses. In this paper, we discuss how we developed an accessibility and inclusion framework, hence a structure for holding ourselves accountable to these principles, for the writing of a statistics textbook. We share our experiences in setting accessibility and inclusion goals, the tools we used to achieve these goals, and recommendations for other educators. We provide examples for instructors that can be implemented in their own courses.
\end{abstract}

\noindent%
\vfill

\newpage
\spacingset{1.45} 

\newcommand{\feedbackform}[1]{https://bit.ly/bayesrules-feedback}

\hypersetup{
    colorlinks,
    linkcolor={blue},
    citecolor={blue},
    urlcolor={blue}
}

\hypertarget{introduction}{%
\section{Introduction}\label{introduction}}

According to LinkedIn's U.S. Emerging Jobs Report, data scientists rank among the top emerging jobs with a 37\% annual growth rate \citeyearpar{linkedin}.
The current tech workforce, unfortunately, lacks gender and racial diversity.
Facebook's Diversity Report notes that only 4.3\% of technical roles are held by Hispanic employees, 1.7\% by Black employees and 24.1\% by women \citeyearpar{facebook}. The figures are also low for Google's technical roles where 0.7\% are held by Black women, 1.1\% by Latina women, and 0.2\% by Native American women \citeyearpar{google}.

In addition to people of color and women (and their intersection), people with disabilities are underrepresented in, and sometimes excluded from, the workforce.
For instance, whereas the Centers for Disease Control and Prevention \citeyearpar{cdc} estimate that 26\% of US adults have some type of disability, only 6.1\% of Google's employees (technical and non-technical) report having a disability.
Employment is considered to be the best route to independence for people with disabilities, yet underemployment is a persistent problem, with education level being one barrier to entry.
Within this group, the employment rate of those with bachelor's degrees is estimated to be triple of those with high school diplomas \citep{yeager}.

Not only does the lack of diversity in the workforce prevent equal opportunities for people who are members of groups that have been (and continue to be) underserved and/or excluded from the workforce, it leads to the creation of products and policies that perpetuate inequality \citep{noble}.
For instance, statistical translation tools (e.g.~Google Translate) yield male pronoun defaults more frequently \citep{prates}.
Facial recognition algorithms are much less accurate in identifying women than men, and darker-skinned people than lighter-skinned people \citep{raji}.
Scholars also raise concerns about
the potential misclassification of people as non-humans by autonomous vehicles after death of a pedestrian by autonomous vehicle while pushing her bicycle.
Such misclassification, especially put people in wheel-chairs at risk \citep{whittaker}.

The lack of diversity in the broader STEM workforce is due in part to a ``leaky pipeline'' \citep{preston}, and a good portion of this leak happens at the college level.
To this end, a healthy learning environment and departmental climate can be factors in the persistence of students who are members of groups that have been underserved and excluded from STEM \citep{packard}.
Thus as statisticians who educate the next generation of the workforce, accessibility and inclusion should be of utmost importance in our programs and courses.

Despite its importance, the typical graduate program in statistics does not provide any training on accessibility and inclusion.
Thus as statistics educators, we must proactively seek out and create a comprehensive accessibility and inclusion framework to shape our teaching, hence a structure for holding ourselves accountable to these principles.
We felt this absence in our recent collaboration on writing a statistics textbook.
In this paper, we discuss how we developed a framework for our own book and teaching.
We share the inclusion and accessibility goals that we set for writing our book, the process and the tools we used to reach these goals, and recommendations for other educators.
Though we specifically share experiences from writing our book, learning about and applying accessibility and inclusion principles has been an iterative process for us.
The changes we made to our book influenced our other teaching practices and vice-versa.
We aim to provide readers with examples that can be implemented in their own courses even if they do not plan on writing a book.

\hypertarget{accessibility-and-inclusion-goals}{%
\section{Accessibility and Inclusion Goals}\label{accessibility-and-inclusion-goals}}

We set two broad goals for our book writing and editing process.

\textbf{Goal 1. The book should be physically accessible.}
In general, readers cannot engage with a book that they cannot access.
To this end, we committed early on to offering online, open access to our book.
The reasons are many.
First and foremost, many students can't afford physical textbooks.
It is estimated that 65\% of student consumers opt out of buying a college textbook and, of those students, 94\% state that they suffer academically \citep{uspirg}.
Second, hosting the book online makes it more accessible to readers across the world.
In addition to the accessibility afforded by an open-access text, we prioritized making our book accessible to readers with visual impairments, including color blind and blind readers.

\textbf{Goal 2. The book should be inclusive of a diverse body of learners.}
Our students and readers don't all share the same gender, race, cultural, socioeconomic class, immigration status, age, sexuality, and religious identities.
They have varying interests, academic experiences, and goals.
We wanted \emph{all} students to be seen, supported, and encouraged to engage with our book.
More broadly, we wanted to create a sense of belonging, not only in our book, but in the broader statistics and data science community.
We took several strategies to this end:

\begin{itemize}
\tightlist
\item
  showcasing the diversity of the field through a broad group of scholars;
\item
  utilizing inclusive language, assumptions, and examples;
\item
  explicitly acknowledging and normalizing the difficulties and failures which are \emph{critical} to learning; and
\item
  building rapport by using a welcoming tone and simply being ourselves.
\end{itemize}

\hypertarget{accessibility-and-inclusion-approaches}{%
\section{Accessibility and Inclusion Approaches}\label{accessibility-and-inclusion-approaches}}

Meeting our goals around inclusion and accessibility could not happen without intention and accountability.
Throughout the writing and editing process, we developed and returned to a set of specific strategies.
We introduce those strategies here, along with the tools we utilized to implement these strategies.
Reflecting the software used in the book and our courses, these particular tools are largely based in R.
Yet the broader themes should resonate no matter your preferred software.

\hypertarget{increasing-accessibility}{%
\subsection{Increasing Accessibility}\label{increasing-accessibility}}

\hypertarget{open-access-materials}{%
\subsubsection{Open-Access Materials}\label{open-access-materials}}

In preparing our open-access book we relied on multiple tools that make online book publishing easier than we would have imagined.
We used the R \texttt{bookdown} package \citep{bookdown-R} to write the book, GitHub for collaboration among us, and Netlify to deploy the book online.
This particular workflow is also effective in developing other types of educational materials, including websites, blogs, and course manuals.
Thus the benefits of investing time to learn these tools can extend beyond writing a book.
Though there's a learning curve here, we found the following resources to be particularly helpful.

\begin{itemize}
\tightlist
\item
  For those who already use R Markdown, the documentation for \textbf{bookdown} is quite extensive \citep{bookdown}. For those who are not familiar with R Markdown, RStudio's online tutorials can provide a starting point \citep{rmarkdown-tutorial}.
\item
  \citet{bryan2021} is an excellent resource for learning \textbf{GitHub} and version control.
\item
  Hvitfeldt's blog post \citeyearpar{emil} provides step-by-step instructions for deploying online materials with \textbf{Netlify}.
\end{itemize}

\hypertarget{color-blind-accessibility}{%
\subsubsection{Color Blind Accessibility}\label{color-blind-accessibility}}

Throughout our book, we utilize full color figures to visualize data on different groups and to compare different models.
Interpreting and learning from these figures thus requires the distinction between multiple colors.
As such, it's critical to be mindful of the fact that there are several types of color blindness.
We took two strategies here.
First, we created the majority of figures in our book using the \texttt{ggplot2} package in R.
Yet instead of relying on the default \texttt{ggplot2} color palette, which is \emph{not} accessible to the most common types of color blindness (eg: red-green color blindness), we utilized the color palette suggested by \citet{okabe}.
The \texttt{palette.colors()} function lists the Okabe-Ito color palette in both hex code and R form.

\begin{Shaded}
\begin{Highlighting}[]
\CommentTok{\# Check out the color palette}
\FunctionTok{palette.colors}\NormalTok{(}\AttributeTok{palette =} \StringTok{"Okabe{-}Ito"}\NormalTok{)}
\DocumentationTok{\#\#         black        orange       skyblue   bluishgreen        yellow }
\DocumentationTok{\#\#     "\#000000"     "\#E69F00"     "\#56B4E9"     "\#009E73"     "\#F0E442" }
\DocumentationTok{\#\#          blue    vermillion reddishpurple          gray }
\DocumentationTok{\#\#     "\#0072B2"     "\#D55E00"     "\#CC79A7"     "\#999999"}
\end{Highlighting}
\end{Shaded}

\noindent Further, placing the following code at the front of the book sets this to be the default color palette throughout:

\begin{Shaded}
\begin{Highlighting}[]
\CommentTok{\# Set the ggplot2 color palette }
\FunctionTok{palette}\NormalTok{(}\StringTok{"Okabe{-}Ito"}\NormalTok{)}
\NormalTok{scale\_colour\_discrete }\OtherTok{\textless{}{-}} \ControlFlowTok{function}\NormalTok{(...) }
  \FunctionTok{scale\_colour\_manual}\NormalTok{(}\AttributeTok{values =} \FunctionTok{palette}\NormalTok{())}
\NormalTok{scale\_fill\_discrete }\OtherTok{\textless{}{-}} \ControlFlowTok{function}\NormalTok{(...) }
  \FunctionTok{scale\_fill\_manual}\NormalTok{(}\AttributeTok{values =} \FunctionTok{palette}\NormalTok{())}
\end{Highlighting}
\end{Shaded}

\begin{figure}

{\centering \includegraphics[width=1\linewidth]{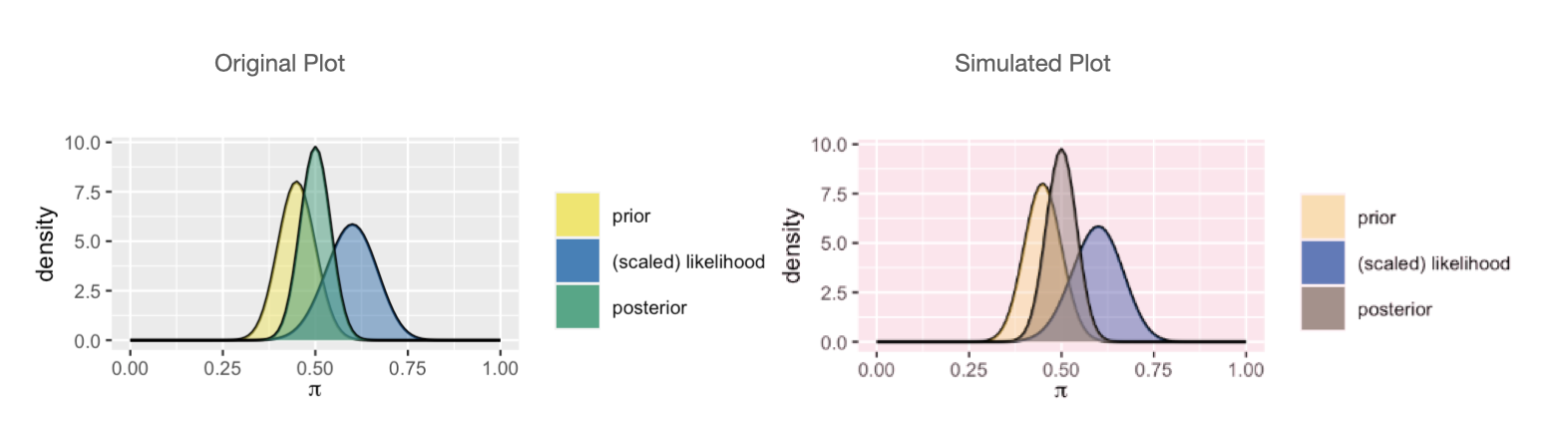} 

}

\caption{A comparison of plots with original color and simulated color.}\label{fig:simulator}
\end{figure}

In addition to largely relying on colors within the Okabe-Ito palette, we tested our color choices with a color blindness simulator \citep{coblis}.
Color blindness simulators help those that don't have any color vision deficiencies see how an image would look to people that do.
For instance, Figure \ref{fig:simulator} displays an original plot that uses our color palette of choice (left) alongside a simulated version (right) of how this plot would look to someone who is green-blind, i.e.~has deuteranopia.
In this case, we learn that some readers will not distinguish the colors here as ``yellow'', ``green'', and ``blue'' (thus we shouldn't rely on those descriptions in our discussion), yet they will view these colors as distinct.

\hypertarget{screen-reader-accessibility}{%
\subsubsection{Screen Reader Accessibility}\label{screen-reader-accessibility}}

In the process of making the book accessible to blind readers, we ran into obstacles, made mistakes, and, to the best of our knowledge, resolved them.
We share this learning process, as well as our final approach.

Our first step in self-teaching on visual accessibility was to download a \textbf{screen reader}.
Screen reader software programs read the content on a computer screen out loud.
We highly encourage non-blind educators to test out a screen reader at least once to experience how inaccessible web resources can be.
In general, screen readers are effective at reading basic text and code, yet can fail to process another major component of the typical statistics and data science resource: figures.
Unless a figure has accompanying \textbf{alt (or alternative) text}, it is read simply as an ``image'' by a screen reader.
As such, alt text which describes the contents of an image for screen readers is an important supplement to figure captions and discussions which appear within the text.

\begin{figure}

{\centering \includegraphics[width=0.5\linewidth]{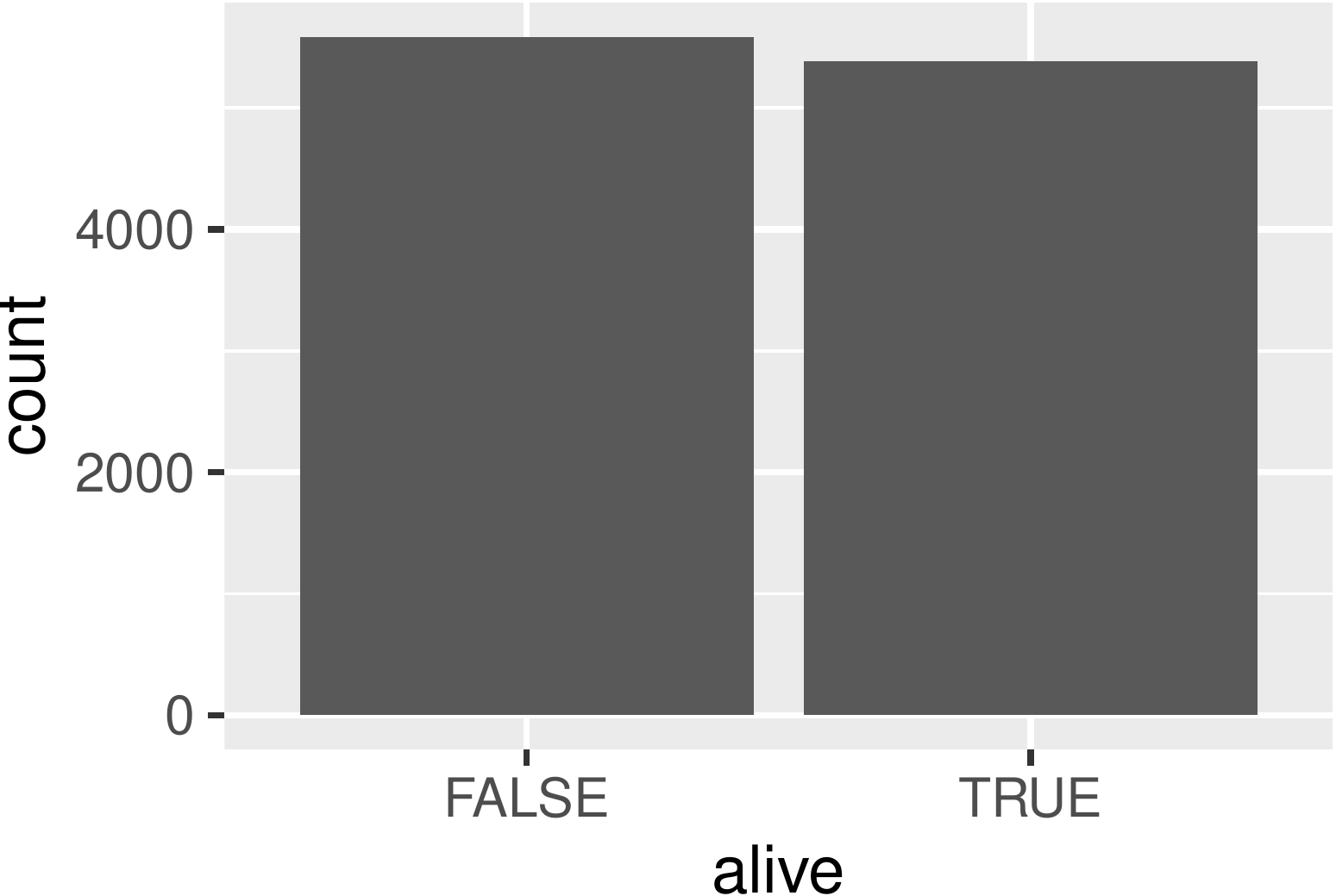} 

}

\caption{The number of living (TRUE) and non-living (FALSE) artists in the MoMA collection.}\label{fig:rmdexample}
\end{figure}

At the onset of our writing process, the \texttt{knitr} \citep{knitr} package behind R Markdown files did not yet support alt text.
Thus we initially considered two primitive approaches to providing alt text for the figures our book: including alt text as comments in our R Markdown source files and providing alt texts in a separate file for download.
Fortunately, as of version 1.32, \texttt{knitr} supports alt text \citep{hill}.\footnote{The authors requested this feature through a GitHub issue.}
Consider the bar plot in Figure \ref{fig:rmdexample} which displays the number of artists in the Museum of Modern Art (MoMA) collection that are still living (TRUE) versus those that aren't (FALSE).
We created this figure, along with alt text, using the code below.
Whereas the figure \emph{caption} (\texttt{fig.cap}) provides a short description of the image which appears in the text \emph{and} is read by screen readers, the alt text (\texttt{fig.alt}) is only read by screen readers.

\begin{Shaded}
\begin{Highlighting}[]
\InformationTok{\textasciigrave{}\textasciigrave{}\textasciigrave{}\{r, fig.cap = "The number of living (TRUE) and non{-}living (FALSE) }
\InformationTok{artists in the MoMA collection.", fig.alt = "The bar plot displays the }
\InformationTok{alive variable on the x{-}axis with FALSE and TRUE values and counts on the}
\InformationTok{y{-}axis. Both the FALSE and TRUE categories have more than 5000 counts but}
\InformationTok{the FALSE category has a slightly higher count than the TRUE category."\}}

\InformationTok{library(bayesrules)}
\InformationTok{library(ggplot2)}

\InformationTok{example\_barplot \textless{}{-} ggplot(moma, aes(x = alive)) +}
\InformationTok{  geom\_bar()}
\InformationTok{example\_barplot }
\end{Highlighting}
\end{Shaded}

In addition to writing our own alt text using the functionality of the \texttt{knitr} package, we learned of the \texttt{BrailleR} \citep{brailler} package which provides \emph{automated} alt text.
For example, the \texttt{VI()} function in \texttt{BrailleR} automates the following alt text for Figure \ref{fig:rmdexample}.

\begin{Shaded}
\begin{Highlighting}[]
\NormalTok{BrailleR}\SpecialCharTok{::}\FunctionTok{VI}\NormalTok{(example\_barplot) }
\DocumentationTok{\#\# This is an untitled chart with no subtitle or caption.}
\DocumentationTok{\#\# It has x{-}axis \textquotesingle{}alive\textquotesingle{} with labels FALSE and TRUE.}
\DocumentationTok{\#\# It has y{-}axis \textquotesingle{}count\textquotesingle{} with labels 0, 2000 and 4000.}
\DocumentationTok{\#\# The chart is a bar chart with 2 vertical bars.}
\DocumentationTok{\#\# Bar 1 is centered horizontally at FALSE, and spans vertically from 0 to 5584.}
\DocumentationTok{\#\# Bar 2 is centered horizontally at TRUE, and spans vertically from 0 to 5380.}
\end{Highlighting}
\end{Shaded}

Resources like \texttt{BrailleR} provide an accessible and efficient approach to including alt text.
Relying on automated alt text is especially appealing in large projects like ours which includes \emph{hundreds} of figures.
Yet the current toolbox for automated alt text has some drawbacks.
First, though it can be effective for simple plots like that in Figure \ref{fig:rmdexample}, automated alt text does not always adequately summarize the nuances and bigger themes within a figure.
Second, automated alt text tools are not yet compatible with more advanced plotting options (e.g.~the Markov chain trace plots within our book).
Thus we are currently revising our alt texts and expect them to evolve along with our understanding of effective practices.
Throughout this revision process, we have found the talk by \citet{canelon} and the supplementary resources in the GitHub repo \citeyearpar{canelonrepo} to be particularly helpful.

Finally, in addition to providing alt text, we improved the screen reader experience by removing the hashtags in R output.
By default, R displays output with double hashtags:

\begin{Shaded}
\begin{Highlighting}[]
\FunctionTok{mean}\NormalTok{(}\FunctionTok{c}\NormalTok{(}\DecValTok{2}\NormalTok{, }\DecValTok{3}\NormalTok{, }\DecValTok{4}\NormalTok{))}
\DocumentationTok{\#\# [1] 3}
\end{Highlighting}
\end{Shaded}

\noindent These hashtags are then read by screen readers, unnecessarily distracting from the code.
To remove the hashtags throughout the book, we set the global \texttt{comment} option to blank (\texttt{""}) at the start of our document:

\begin{Shaded}
\begin{Highlighting}[]
\NormalTok{knitr}\SpecialCharTok{::}\NormalTok{opts\_chunk}\SpecialCharTok{$}\FunctionTok{set}\NormalTok{(}\AttributeTok{comment =} \StringTok{""}\NormalTok{)}
\end{Highlighting}
\end{Shaded}

\noindent The result applied to our mean calculation is illustrated below:

\begin{Shaded}
\begin{Highlighting}[]
\FunctionTok{mean}\NormalTok{(}\FunctionTok{c}\NormalTok{(}\DecValTok{2}\NormalTok{, }\DecValTok{3}\NormalTok{, }\DecValTok{4}\NormalTok{))}
\NormalTok{[}\DecValTok{1}\NormalTok{] }\DecValTok{3}
\end{Highlighting}
\end{Shaded}

\hypertarget{increasing-inclusivity}{%
\subsection{Increasing Inclusivity}\label{increasing-inclusivity}}

\hypertarget{showcasing-the-diversity-within-statistics-and-data-science}{%
\subsubsection{Showcasing the Diversity within Statistics and Data Science}\label{showcasing-the-diversity-within-statistics-and-data-science}}

One simple strategy for making our book more inclusive of \emph{students} with different identities and experiences is to make it more inclusive of \emph{scholars} in the fields of statistics and data science.
The Data Feminism book \citep{datafeminism1} provided a source of inspiration here.
At the onset of their project, the authors outlined a set of values and aspirational metrics for holding themselves accountable to these values \citep{datafeminism}.
For example, to address structural racism, the authors aspired to have at least 75\% of their citations of feminist scholarship to be from people of color.

Though we did not set any numerical diversity metrics, we were mindful of the scholars we cited and presented in our book.
If we were \emph{not} intentional about citing a diverse body of scholars, we feared that we would perpetuate the visibility of a narrow, homogeneous, and extensively cited group at the expense of others that are also doing groundbreaking work.
Thus throughout the writing and revision process, we reflected on whether the cited scholars represented diverse fields and demographics.
To learn more about the works of underrepresented minority scholars ourselves, the following organizations provided invaluable resources: Mathematically Gifted and Black, the Society for Advancement of Chicanos/Hispanics and Native Americans in Science, the American Statistical Association (ASA) Committee on Minorities in Statistics, and the ASA Committee on Women in Statistics.
In addition, we believe that the ASA's new Justice, Equity, Diversity, and Inclusion Outreach group (JEDI) will further inform our diversity goals in both writing and teaching.

Not only did we want to be mindful of \emph{who} we cite, but \emph{how} we cite.
For example, trans scholars might have historical publications under their deadnames since many publishers make it difficult to correct their names \citep{tanenbaum}.
Further, we cannot infer a scholar's gender identity by their name alone.
Though it is difficult to assess the appropriate names and gender pronouns for deceased scholars, we relied on personal websites, ORCID IDs, and social media accounts to help determine this information for living scholars.
We intentionally avoided using publications and Google scholar for names and pronouns.

\hypertarget{utilizing-inclusive-language-assumptions-and-examples}{%
\subsubsection{Utilizing Inclusive Language, Assumptions, and Examples}\label{utilizing-inclusive-language-assumptions-and-examples}}

Throughout our book, we aimed for our language, assumptions, and examples to be mindful of two facts: not all readers share the same specialized knowledge; and not all readers share the same cultural or personal identities and lived experiences.
Consider the first of these.
Statistical applications often require deep, specialized knowledge (e.g.~about biology, physics, sports, etc).
Assuming that all readers share the same specialized knowledge is problematic for a couple of reasons.
First, trying to teach a technical statistical concept through a specialized context is like trying to teach kids to read using passages about retirement plans -- it's challenging if not impossible to learn a concept when it's delivered within a context that you don't understand.
Second, specialized settings risk alienating readers which, in turn, can detract from the learning process.
To this end, we aimed to showcase applications that are either (fairly) universal or at least have a well explained context.
For example, we paid close attention to the fact that we teach many international students and that we expect to have readers outside the U.S..
As such, we eliminated one of our original examples which required an understanding of the unique primary and caucus phase of U.S. election cycles.
Similarly, we eliminated an example about two popular U.S. restaurants, Starbucks and Dunkin', realizing that it may not be relevant for many.
Though we authors bring a variety of perspectives and experiences to our book, we also acknowledge that evaluating our own work along these lines, and recognizing our own niche knowledge bases, can be challenging.

Next, we aimed for our book to reflect the fact that not all readers share the same cultural or personal identities and lived experiences.
As one example, we avoided cultural stereotypes and used names that are associated with different cultures.
In addition to ``Matt'' and ``Taylor'', there's Muhammad, Zuofu, Kimya, and Fernando.
We also paid acute attention to the evolving and more inclusive understanding of sexual orientation and gender identity \citep{phillips2019}.
To this end, we employed three strategies to avoid heteronormative defaults.
First, our examples span topics from same-sex marriage to LGBTQ+ anti-discrimination laws.
Second, we use three singular person pronouns, he / she / they, when referring to people in our examples.
Third, as recommended by \citet{thornton2019}, we avoided examples and datasets that treat gender as binary.
Given the lack of datasets with a non-binary gender variable, this means that we largely avoided gender altogether, yet developed some of our own hypothetical examples.
For example, we include an exercise in which a person meets men, women, non-binary people, and people with other gender identities on a dating app.

\hypertarget{embracing-difficulties-and-failures}{%
\subsubsection{Embracing Difficulties and Failures}\label{embracing-difficulties-and-failures}}

Whether in the classroom or a book, assumptions and statements about what should be ``obvious'' to our students can be both wrong and demoralizing.
Throughout our book, we worked to actively embrace different academic backgrounds as well as the mistakes and questions that are critical to learning.
For example, we explicitly state up front that: ``As you read the book and put Bayesian methodology into practice, you will make mistakes. Many mistakes. Making and learning from mistakes is simply part of learning.''
As another example, we note in the second chapter of the book that, no matter their level of previous experience, readers will be introduced to many new vocabulary terms and concepts, and so should expect to take this chapter slowly.
The hope here is to counter any impostor syndrome by normalizing the fact that humans don't magically ingest and process information -- learning takes time and iteration.
Finally, we aimed to use language that creates a welcoming environment for readers with a variety of technical or academic backgrounds.
For example, we don't assume that all readers know the Greek alphabet, thus give readers a heads up on Greek letters and how to pronounce them.

\hypertarget{building-rapport}{%
\subsubsection{Building Rapport}\label{building-rapport}}

Our final strategy for creating an inclusive resource was to build rapport.
Since we're not in the same room together, building rapport between authors and readers is a bit tougher than between a teacher and students in the same classroom.
Yet it's still important and possible.
Our simple approach here was really just to be ourselves, not textbook writing robots.
For example, our book logo is a somewhat silly disco ball, a loose analogy of the Bayesian philosophy.
We use casual imagery and analogies to build intuition for technical concepts.
Similarly, our discussions tend to be conversational in tone.
In general, our hope is that by using informal language to bring to life formal concepts, and by sharing some of our own personalities, readers feel more seen themselves.

\hypertarget{concluding-remarks}{%
\section{Concluding Remarks}\label{concluding-remarks}}

Table \ref{tab:checklist} summarizes our strategies to be intentional about, and hold ourselves accountable to, our accessibility and inclusion goals.
Though we originally developed these strategies for use in the writing and revision of our book, we have since utilized them in our classrooms.
It is possible that someone reading our book might disagree that we met all of these criteria.
With the subjective measures as well as our own personal identities and experiences, we almost certainly missing goals without realizing it.
As educators, we also expect our understanding of accessibility and inclusion to evolve over time as we seek to continually improve our teaching practices.
This checklist will become outdated or insufficient quite soon.
With this understanding, we consider the framework we've provided here to be a \emph{starting point} for continued collaboration and conversation with other statistics and data science educators.
As our understanding and practice around accessibility and inclusion evolve we welcome any feedback on these topics through a Google form\footnote{The Google form can be found at \url{\feedbackform{}}}.

\begin{table}[!h]

\caption{\label{tab:checklist}Checklist for Evaluating Accessibility and Inclusion of Teaching Materials}
\centering
\begin{tabular}[t]{>{\raggedright\arraybackslash}p{10em}>{\raggedright\arraybackslash}p{25em}}
\toprule
Accessibility and Inclusion Criteria & Questions\\
\midrule
Acessibility & Is the cost affordable for learners from diverse socioeconomic backgrounds?\\
 & Are plots distinguishable to color blind learners?\\
 & Is alt text provided for images?\\
\midrule
Inclusivity of scholars & Do the cited scholars represent diversity across identities, experiences, and expertise?\\
 & Are scholars cited using the correct names and pronouns?\\
\midrule
\addlinespace
Inclusivity of students & Do examples avoid the necessity of specialized knowledge?\\
 & Do names and pronouns reflect diverse cultural and personal identities?\\
 & Are there examples that could potentially speak to younger as well as older students?\\
 & Does the delivery embrace mistakes and critical thinking?\\
 & Are efforts made to accommodate different academic experiences and create a shared foundation?\\
\bottomrule
\end{tabular}
\end{table}

\newpage

\hypertarget{appendix}{%
\section*{Appendix}\label{appendix}}
\addcontentsline{toc}{section}{Appendix}

We provide the R session information used writing this manuscript.

\begin{Shaded}
\begin{Highlighting}[]
\FunctionTok{sessionInfo}\NormalTok{()}
\DocumentationTok{\#\# R version 4.1.0 (2021{-}05{-}18)}
\DocumentationTok{\#\# Platform: x86\_64{-}apple{-}darwin17.0 (64{-}bit)}
\DocumentationTok{\#\# Running under: macOS Big Sur 10.16}
\DocumentationTok{\#\# }
\DocumentationTok{\#\# Matrix products: default}
\DocumentationTok{\#\# BLAS:   /Library/Frameworks/R.framework/Versions/4.1/Resources/lib/libRblas.dylib}
\DocumentationTok{\#\# LAPACK: /Library/Frameworks/R.framework/Versions/4.1/Resources/lib/libRlapack.dylib}
\DocumentationTok{\#\# }
\DocumentationTok{\#\# locale:}
\DocumentationTok{\#\# [1] en\_US.UTF{-}8/en\_US.UTF{-}8/en\_US.UTF{-}8/C/en\_US.UTF{-}8/en\_US.UTF{-}8}
\DocumentationTok{\#\# }
\DocumentationTok{\#\# attached base packages:}
\DocumentationTok{\#\# [1] stats     graphics  grDevices utils     datasets  methods   base     }
\DocumentationTok{\#\# }
\DocumentationTok{\#\# other attached packages:}
\DocumentationTok{\#\# [1] ggplot2\_3.3.5         bayesrules\_0.0.2.9000 kableExtra\_1.3.4     }
\DocumentationTok{\#\# }
\DocumentationTok{\#\# loaded via a namespace (and not attached):}
\DocumentationTok{\#\#   [1] readxl\_1.3.1         systemfonts\_1.0.2    plyr\_1.8.6          }
\DocumentationTok{\#\#   [4] igraph\_1.2.6         splines\_4.1.0        crosstalk\_1.1.1     }
\DocumentationTok{\#\#   [7] usethis\_2.0.1        rstantools\_2.1.1     inline\_0.3.19       }
\DocumentationTok{\#\#  [10] digest\_0.6.28        htmltools\_0.5.2      rsconnect\_0.8.24    }
\DocumentationTok{\#\#  [13] fansi\_0.5.0          magrittr\_2.0.1       memoise\_2.0.0       }
\DocumentationTok{\#\#  [16] remotes\_2.4.0        extrafont\_0.17       RcppParallel\_5.1.4  }
\DocumentationTok{\#\#  [19] matrixStats\_0.61.0   xts\_0.12.1           extrafontdb\_1.0     }
\DocumentationTok{\#\#  [22] svglite\_2.0.0        prettyunits\_1.1.1    colorspace\_2.0{-}2    }
\DocumentationTok{\#\#  [25] rvest\_1.0.1          xfun\_0.26            dplyr\_1.0.7         }
\DocumentationTok{\#\#  [28] callr\_3.7.0          crayon\_1.4.1         jsonlite\_1.7.2      }
\DocumentationTok{\#\#  [31] roloc\_0.1{-}1          lme4\_1.1{-}27.1        rolocISCCNBS\_0.1    }
\DocumentationTok{\#\#  [34] survival\_3.2{-}13      zoo\_1.8{-}9            glue\_1.4.2          }
\DocumentationTok{\#\#  [37] gtable\_0.3.0         webshot\_0.5.2        hunspell\_3.0.1      }
\DocumentationTok{\#\#  [40] V8\_3.4.2             pkgbuild\_1.2.0       Rttf2pt1\_1.3.9      }
\DocumentationTok{\#\#  [43] rstan\_2.21.2         scales\_1.1.1         DBI\_1.1.1           }
\DocumentationTok{\#\#  [46] miniUI\_0.1.1.1       Rcpp\_1.0.7           groupdata2\_1.5.0    }
\DocumentationTok{\#\#  [49] viridisLite\_0.4.0    xtable\_1.8{-}4         reticulate\_1.22     }
\DocumentationTok{\#\#  [52] gridGraphics\_0.5{-}1   proxy\_0.4{-}26         stats4\_4.1.0        }
\DocumentationTok{\#\#  [55] StanHeaders\_2.21.0{-}7 DT\_0.19              htmlwidgets\_1.5.4   }
\DocumentationTok{\#\#  [58] httr\_1.4.2           threejs\_0.3.3        ellipsis\_0.3.2      }
\DocumentationTok{\#\#  [61] pkgconfig\_2.0.3      loo\_2.4.1            XML\_3.99{-}0.8        }
\DocumentationTok{\#\#  [64] farver\_2.1.0         utf8\_1.2.2           janitor\_2.1.0       }
\DocumentationTok{\#\#  [67] tidyselect\_1.1.1     labeling\_0.4.2       rlang\_0.4.11        }
\DocumentationTok{\#\#  [70] reshape2\_1.4.4       later\_1.3.0          cellranger\_1.1.0    }
\DocumentationTok{\#\#  [73] munsell\_0.5.0        tools\_4.1.0          cachem\_1.0.6        }
\DocumentationTok{\#\#  [76] cli\_3.0.1            moments\_0.14         generics\_0.1.0      }
\DocumentationTok{\#\#  [79] devtools\_2.4.2       ggridges\_0.5.3       evaluate\_0.14       }
\DocumentationTok{\#\#  [82] stringr\_1.4.0        fastmap\_1.1.0        yaml\_2.2.1          }
\DocumentationTok{\#\#  [85] rticles\_0.21         processx\_3.5.2       knitr\_1.35          }
\DocumentationTok{\#\#  [88] fs\_1.5.0             purrr\_0.3.4          nlme\_3.1{-}153        }
\DocumentationTok{\#\#  [91] whisker\_0.4          mime\_0.11            rstanarm\_2.21.1     }
\DocumentationTok{\#\#  [94] xml2\_1.3.2           compiler\_4.1.0       bayesplot\_1.8.1     }
\DocumentationTok{\#\#  [97] shinythemes\_1.2.0    rstudioapi\_0.13      png\_0.1{-}7           }
\DocumentationTok{\#\# [100] curl\_4.3.2           BrailleR\_0.32.0      e1071\_1.7{-}9         }
\DocumentationTok{\#\# [103] testthat\_3.0.4       tibble\_3.1.4         stringi\_1.7.4       }
\DocumentationTok{\#\# [106] ps\_1.6.0             desc\_1.4.0           lattice\_0.20{-}45     }
\DocumentationTok{\#\# [109] Matrix\_1.3{-}4         nloptr\_1.2.2.2       markdown\_1.1        }
\DocumentationTok{\#\# [112] shinyjs\_2.0.0        vctrs\_0.3.8          pillar\_1.6.3        }
\DocumentationTok{\#\# [115] lifecycle\_1.0.1      httpuv\_1.6.3         R6\_2.5.1            }
\DocumentationTok{\#\# [118] bookdown\_0.24        promises\_1.2.0.1     gridSVG\_1.7{-}2       }
\DocumentationTok{\#\# [121] gridExtra\_2.3        sessioninfo\_1.1.1    codetools\_0.2{-}18    }
\DocumentationTok{\#\# [124] boot\_1.3{-}28          colourpicker\_1.1.0   MASS\_7.3{-}54         }
\DocumentationTok{\#\# [127] gtools\_3.9.2         assertthat\_0.2.1     pkgload\_1.2.2       }
\DocumentationTok{\#\# [130] rprojroot\_2.0.2      nortest\_1.0{-}4        withr\_2.4.2         }
\DocumentationTok{\#\# [133] shinystan\_2.5.0      parallel\_4.1.0       grid\_4.1.0          }
\DocumentationTok{\#\# [136] class\_7.3{-}19         minqa\_1.2.4          rmarkdown\_2.11      }
\DocumentationTok{\#\# [139] snakecase\_0.11.0     shiny\_1.7.0          lubridate\_1.7.10    }
\DocumentationTok{\#\# [142] base64enc\_0.1{-}3      dygraphs\_1.1.1.6}
\end{Highlighting}
\end{Shaded}

\newpage

\bibliographystyle{agsm}
\bibliography{bibliography}

\end{document}